\documentclass[12pt]{iopart}
\usepackage{iopams}
\usepackage{graphics}
\usepackage{graphicx}
\usepackage{bm}
\usepackage{amsfonts}
\usepackage{amssymb}
\usepackage{latexsym}
\usepackage{color}
\usepackage{bbold}
\usepackage{braket}
\usepackage{hyperref}
\usepackage{enumitem}

\newcommand{\beq}{\begin{eqnarray}}
\newcommand{\eeq}{\end{eqnarray}}
\newcommand{\eq}[1]{Eq.~(\ref{#1})}
\newcommand{\fig}[1]{Fig.~\ref{#1}}

\usepackage{ulem}

\begin{document}

\title[Mitigating decoherence in hot electron interferometry]{Mitigating decoherence in hot electron interferometry}

\author{Lewis A. Clark$^1$, Masaya Kataoka$^2$, Clive Emary$^1$}

\address{$^1$ Joint Quantum Centre Durham-Newcastle, School of Mathematics, Statistics and Physics, Newcastle University, Newcastle upon Tyne NE1 7RU, United Kingdom}
\address{$^2$ National Physical Laboratory, Hampton Road, Teddington, Middlesex TW11 0LW, United Kingdom}

\begin{abstract}
	Due to their high energy, hot electrons in quantum Hall edge states can be considered as single particles that have the potential to be used for quantum optics-like experiments.  Unlike photons, however, electrons typically undergo scattering processes in transport, which results in a loss of coherence and limits their ability to show quantum-coherent behaviour.  
	Here we study theoretically the decoherence mechanisms of hot electrons in a Mach-Zehnder interferometer, and highlight the role played by both acoustic and optical phonon emission. We discuss optimal choices of experimental parameters and
	show that high visibilities of $\gtrsim 85\%$ are achievable in hot-electron devices over relatively long distances of 10 $\mu$m. 
	We also discuss energy filtration techniques to remove decoherent electrons and show that this can increase  visibilities to over $95\%$. This represents an improvement over Fermi-level electron quantum optics, and suggests hot-electron charge pumps as a platform for realising quantum-coherent nanoelectronic devices.
\end{abstract}

\vspace{10pt}
\begin{indented}
	\item[]August 2020
\end{indented}

\section{Introduction}
The realisation of quantum optics experiments with electrons is a long-standing pursuit of the mesoscopic physics community \cite{EQO_review,EQO_review2}.  In a typical architecture, quantum Hall edge (QHE) channels form the electronic analogue of photonic waveguides and quantum point contacts (QPCs) act as beamsplitters.  A number of classic optical experiments have been realised in this way: Hanbury-Brown-Twiss \cite{eHBT} and Hong-Ou-Mandel \cite{eHOM} experiments; as well as the focus of this work, the Mach-Zehnder interferometer (MZI) \cite{eMZI,eMZI_exp,eMZI_decoherence2,eMZI_exp2,Haack}.

One significant difference between electrons and photons is that electrons interact strongly with their environment.  Early electron interferometers demonstrated relatively low visibilities \cite{eMZI,eMZI_exp,eMZI_decoherence2,eMZI_exp2,Haack}, up to $\sim 80\%$ \cite{exp_vis} for arm lengths of 8 $\mu$m, the cause of which has been investigated in detail \cite{eMZI_decoherence,eMZI_decoherence3,Levkivskyi,eMZI_decoherence4,Ferraro_decoherence}.  In particular, the Coulomb interaction via the creation of plasmonic excitations \cite{das_Sarma,Gasser,das_Sarma2,Giuliani_book}, has been highlighted as the dominant relaxation and decoherence mechanism for electrons close to the Fermi surface \cite{Levkivskyi,eMZI_decoherence4}.  In this ``cold-electron'' regime with electrons injected by conventional sources, potential applications might be limited due to the relatively short coherence lengths.

Recent advances in electron pumps \cite{Kaestner2015}, however, have led to technology able to inject single electrons into edge channels at energies well above the Fermi sea  \cite{Charge_pump,Charge_pump2,Charge_pump3,Charge_pump4,Ampere,Charge_pump5,Charge_pump6,Charge_pump7,Charge_pump8} with both a high rate and accuracy. 
These ``hot electrons'' are well separated, both energetically and spatially, from those in the bulk, and it has been suggested that this leads to a significant reduction in the Coulomb scattering of injected electrons \cite{Time_of_flight}.
Indeed, in Ref.~\cite{Phonon_Fujisawa} it was observed that, above a certain energy, electron transport was close to ballistic, indicating a suppression of Coulomb interactions.
Within this new hot-electron regime, however, phonon emission becomes a significant relaxation process, as has been discussed theoretically \cite{LO_theory,acoustic_phonon_theory} and observed experimentally \cite{Phonon_Fujisawa,LO_exp}.  This is potentially a problem because the emission of a phonon reveals ``which-way'' information for the electron, thus destroying its quantum coherence.

\begin{figure}[t] 
	\centering
	\includegraphics[width=0.6\linewidth]{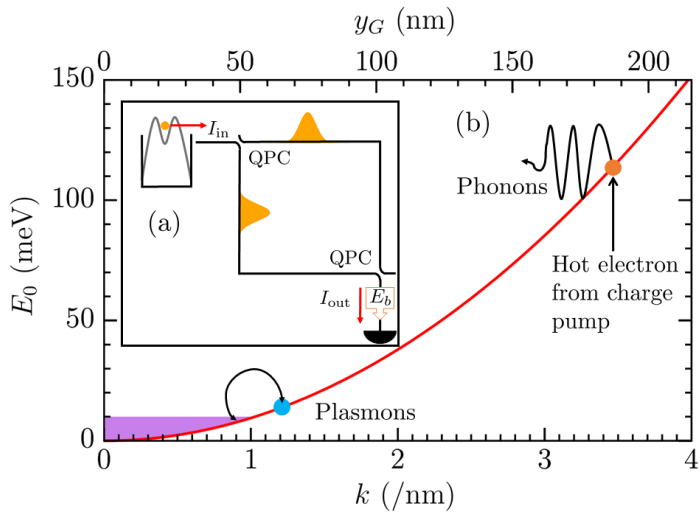}
	\caption{ 
		(a) Schematic hot-electron MZI with a charge pump source and an electron wave packet moving in a QHE channels that is split and recombined by QPC beamsplitters.  Before arriving at the detector, the current $I_{\rm out}$ may be partially blocked by a potential barrier of height $E_b$.  This energy filtration removes decohered electrons and enhances visibility.  
		(b) Dispersion relation of outermost QHE-channel electrons with indication of dominant relaxation processes.  At low injection energy, $E_0$, electrons close to the Fermi sea relax mostly through plasmon creation.  At higher energies, the electron-electron interactions are suppressed due to the large separation between guide centre $y_G$ (transverse average position) and the Fermi sea.  At these high energies, phonon emission becomes the dominant decay mechanism.
	}
	\label{Fig:HEMZI}
\end{figure}

In this paper, we calculate relaxation and decoherence rates of hot electrons in a MZI geometry, and find the dependence of these rates on parameters such as injection energy and magnetic field. This provides a theoretical underpinning to the suppression of electron-electron interactions and the dominance of phonon channels for hot electrons.  
Moreover, we show how a careful choice of experimental parameters can minimise decoherence and maximise interferometer visibility. We find that, without further manipulation of the output signal, visibilities $\gtrsim85\%$ can be obtained in a hot-electron interferometer with arm lengths of $10 \mu$m.
We then go on to consider the effect of introducing a potential barrier before the detection of the output current.  This serves as an energy filter of the MZI signal.  Due to the energetic distinguishability of coherent and incoherent parts of the signal, which although not complete is well expressed in the hot-electron case, this filtration allows the extraction of the coherent contribution of the signal.  By so doing, we find that visibilities of $\gtrsim95\%$ are attainable with arm lengths of 10 $\mu$m, thus achieving very high coherence even over relatively long distances.
In minimising the decoherence and employing the filtration techniques that we outline here, a highly coherent signal of single electrons is produced, suggesting a path towards scalable coherent nanoelectronic devices with single-electron sources.

\section{Model} \label{Sec:Model}

The MZI setup we consider here is sketched in Fig.~\ref{Fig:HEMZI}a.  A dynamic-quantum-dot charge pump injects electrons into a two-dimensional electron gas in the quantum Hall regime.  Injection is into the outermost edge-channel \cite{acoustic_phonon_theory}, and the resultant current of single electrons is split and recombined with QPCs \cite{eMZI}. After this, the current is measured past a potential barrier imposed on the output channel with a static top gate. 
In this paper we consider electrons confined in QHE states with strong confinement in the $z$-direction, weak confinement in the $y$-direction and
transport in the $x$ direction.  We assume the weak edge confinement to be harmonic with frequency $\omega_y$.
In this study, we take $\hbar\omega_y = 2.7$\,meV \cite{Time_of_flight}, but consider how a change in this parameter alters our results later in Sec.~\ref{Sec:Enhancements}.  With this potential, the dispersion relation of the edge-channel electrons is as shown in Fig.~\ref{Fig:HEMZI}b. As indicated, the injection energy of the electrons from the charge pump is high above the Fermi energy.

The relevant quantum numbers here are the wavenumber, $k$, in transport direction, Landau level $m$ and quantum number $n$ describing the $z$-direction subband.
The Hamiltonian describing the single particle states is
\begin{eqnarray}
H_e & = & \sum\limits_{n m k} E_{n m k} c_{n m k}^\dagger c_{n m k} \, ,
\end{eqnarray}
with energies, measured relative to the bottom of the lowest subband, given by
\begin{eqnarray}
E_{n m k} & = & \epsilon_n + m \hbar \Omega + \frac{1}{2} \left(\frac{\Omega}{\omega_c}\right)^2 m_e^* \omega_y^2 y_G^2 (k) \, ,
\end{eqnarray}
with effective electron mass $m_e^* = 0.067 m_e$ \cite{Datta}.
The energy $\epsilon_n$ is from the energy level in the $z$-direction with $\epsilon_1=0$, while $\Omega$ is the effective confinement frequency given by $\Omega = \sqrt{\omega_y^2 + \omega_c^2}$, with cyclotron frequency $\omega_c = \left|e B / m_e^* \right|$, with $e$ the elementary charge.  The guide centre $y_G (k)$ describes the mean $y$ co-ordinate of a confined electron and is given by
\begin{eqnarray}
y_G (k) & = & \frac{\hbar k}{e B} \left(\frac{\omega_c}{\Omega} \right)^2 \, .
\end{eqnarray}
Within this study, we consider only electrons in the first ($n=1,m=0$) and second Landau levels ($n=1, m=1$), as most of the dynamics are expected to occur here \cite{acoustic_phonon_theory}.  The wavefunction describing these states is \cite{Datta}
\begin{eqnarray}
\ket{\Psi_{\bf n}} & = & {\rm e}^{{\rm i} k x} \chi_{m k} (y) \phi (z) \, ,
\end{eqnarray}
with ${\bf n} = (1,m,k)$ and
\begin{eqnarray}
\chi_{m k} (y) & = & \sqrt{\frac{1}{l_\Omega}} \frac{1}{\sqrt{2^m m! \pi^{\frac{1}{4}}}} \exp \left( - \frac{\left(y - y_G(k) \right)^2}{2 l_\Omega^2} \right)  H_m \left(\frac{y-y_G}{l_\Omega} \right) \nonumber \\
\phi(z) & = & \left(2 a^3 \right)^{-1/2} z {\rm e}^{-z/(2a)}\, ,
\end{eqnarray}
where we have assumed a triangular well confinement in the $z$-direction \cite{Tri_well} and $H_m(x)$ is the $m^\mathrm{th}$-order Hermite polynomial.  In this study we consider a width parameter of $a = 3$ nm \cite{acoustic_phonon_theory,LO_exp}, although in Sec.~\ref{Sec:Enhancements} consider how changing this parameter affects our results.

\subsection{Electron-phonon interactions}

The electron may emit a longitudinal-optical (LO) phonon, which are approximately dispersionless with energy 36 meV in GaAs \cite{Taubert}.
The hot electron may also emit longitudinal acoustic phonons via either of the deformation potential (LADP) or piezoelectric field interactions (LAPZ), or transverse phonons via piezoelectric interaction (TAPZ).

To model relaxation due to these processes, we use a Fr\"ohlich Hamiltonian \cite{Frohlich,Mahan_book} which, in the interaction picture, reads
\begin{eqnarray} \label{Eq:Ham_phonon}
V & = & \sum\limits_{{\bf n},{\bf n'},{\bf q},\gamma} \Lambda^{(\gamma)}_{{\bf n} {\bf n'}} c_{{\bf n'}}^\dagger c_{\bf n} \left( {\rm e}^{{\rm i} \left(E_{\bf n'} - E_{\bf n} + \hbar \omega\right)} a_{-{\bf q},\gamma}^\dagger + {\rm e}^{{\rm i} \left(E_{\bf n'} - E_{\bf n} - \hbar \omega\right)} a_{{\bf q},\gamma} \right) \, ,
\end{eqnarray}
where a hot electron is transferred from the QHE state ${\bf n}$ to ${\bf n'}$ and creates/absorbs a phonon of type $\gamma \in \left\{{ \rm LO,LADP,LAPZ,TAPZ}\right\}$, with momentum exchange ${\bf q}$ with matrix element $\Lambda^{(\gamma)}_{{\bf n} {\bf n'}}$ \cite{acoustic_phonon_theory}.

We then extend the above to describe an electron within an interferometer.  We model this situation by assuming that the two arms of the interferometer, with arm index $\alpha = \left\{1,2\right\}$, are separated only in the $y$-direction (perpendicular to the direction of propagation, $x$) by a distance $l_{\rm arm}$, which we will approximate as constant.  In a realistic device, the separation will likely not only be in the $y$-direction but also the $x$-direction, as in Fig.~\ref{Fig:HEMZI}(a).  However, this is a reasonable model to assume, as only for a very small arm separation $l_{\rm arm} \sim l_\Omega$ will the behaviour deviate from this, as we later show.  Such small separations occur only for a very short time in the overall evolution for all realistic devices and as such this approximation holds well.  This results in a phase difference in the phonon field seen by the electron in each of the two arms.  The Hamiltonian then reads
\begin{eqnarray} \label{Eq:Ham}
V & = & \sum\limits_{{\bf n},{\bf n'},{\bf q},\alpha,\gamma} \Lambda^{(\gamma)}_{{\bf n}_\alpha {\bf n'}_\alpha} c_{{\bf n'}_\alpha}^\dagger c_{{\bf n}_{\alpha}} \left( {\rm e}^{{\rm i} \left(E_{{\bf n'}} - E_{\bf n} + \hbar \omega\right)} {\rm e}^{{\rm i} q_y l_{\rm arm} \delta_{\alpha , 2}} a_{-{\bf q},\gamma}^\dagger \right. \nonumber \\
&& 
~~~~~~~~~~~~~~~~~~~~~~~~~~~~~~~~~~
\left. +~ {\rm e}^{{\rm i} \left(E_{\bf n'} - E_{\bf n} - \hbar \omega\right)} {\rm e}^{-{\rm i} q_y l_{\rm arm} \delta_{\alpha , 2}} a_{{\bf q},\gamma} \right) \, .
\end{eqnarray}
In the limiting case of $l_{\rm arm}=0$, we recover the standard Fr\"ohlich Hamiltonian for each arm.
The calculation of the relevant phonon-induced rates using Fermi's golden rule and these Hamiltonians is detailed in \ref{App:Rates}.

\subsection{Electron-electron interactions}

The importance of electron-electron interactions will be assessed by considering the Coulomb Hamiltonian
\begin{eqnarray} \label{Eq:Ham_ee}
H_{e{\rm -}e} & = & \sum\limits_{k,p,q} W_c (q,\omega) \, c_{p-q}^\dagger c_{k+q}^\dagger c_{k} c_{p} \, ,
\end{eqnarray}
for electrons confined within the lowest Landau level (for simplicity we have suppressed the $n=1$ and $m=0$ indices on the electrons operators).
Here, the transfer of electrons with wavevectors $k$ and $p$ to $k+q$ and $p-q$ is mediated by the quasi-1D screened Coulomb potential 
$W_c(q,\omega)  =  \epsilon^{-1} (q,\omega) V_c(q)$\cite{das_Sarma,das_Sarma2,Giuliani_book} with $\epsilon(q,\omega)$ the dielectric function and 
\begin{eqnarray}
V_{c} (q)&=& \int {\rm d} y \int {\rm d} y' \int {\rm d} z \int {\rm d} z' \, v\left(q \sqrt{(y-y')^2 + (z-z')^2}\right) \nonumber \\
&& \times \chi_{0 k} (y) \chi_{0 k+q}^* (y) \chi_{0 p} (y') \chi_{0 p-q}^* (y') \phi (z) \phi^* (z) \phi (z') \phi^* (z')  \, , \nonumber
\end{eqnarray}
where
\begin{eqnarray} \label{Eq:Coulomb2}
v \left( \gamma \right) & = & \frac{e^2}{2 \pi \epsilon_0} K_0 \left( \left| \gamma \right| \right) \, ,
\end{eqnarray}
is the bare quasi-1D Coulomb potential.  The static dielectric constant of the material is denoted $\epsilon_0$, $K_0$ is the zeroth-order modified Bessel function of the second kind and we take $q$ to be the momentum exchange in the $x$-direction.  The frequency $\omega$ corresponds to the energy exchanged in this process.

\section{Relaxation} \label{Sec:Relax}
We first consider relaxation in hot-electron quantum optics as this forms the basis for understanding of decoherence.  Additional details of the calculations in this section are given in \ref{App:Rates}.

\subsection{Phonon emission}
Acoustic phonon emission by electrons in the $m=0$ subband can scatter them not only to the same subband but also into different ones, predominantly with $m=1$. 
As discussed in Refs.~\cite{acoustic_phonon_theory,LO_exp}, relaxation via LO emission from the inner $m=1$ subband back to $m=0$ is typically rapid, with the result being that inter-Landau-level scattering by acoustic phonons can be thought of as giving an enhancement to the $m=0 \to 0$ LO-phonon relaxation rate. Thus we define an ``effective'' rate (labelled LO$_{\rm eff}$) that is the sum of the rates of the direct LO process and that of the two-step process involving the $m=1$ subband:
\begin{eqnarray} \label{Eq:LO_eff}
\Gamma^{\rm LO}_{\rm eff} & = & \Gamma^{\rm LO}_{\rm 00} + \Gamma^{\rm LO}_{\rm 2-step} 
=
\Gamma^{\rm LO}_{\rm 00} + \left(\frac{1}{\Gamma^{\rm LADP}_{10} + \Gamma^{\rm LAPZ}_{10} + \Gamma^{\rm TAPZ}_{10}} + \frac{1}{\Gamma^{\rm LO}_{01}} \right)^{-1} \, ,
\label{EQ:GammaLOeff}
\end{eqnarray}
where $\Gamma^{(\gamma)}_{m' m}$ is the decay rate for phonon type $\gamma$ going from Landau level $m$ to $m'$. 

\begin{figure}[t] 
	\centering
	\includegraphics[width=0.99\textwidth]{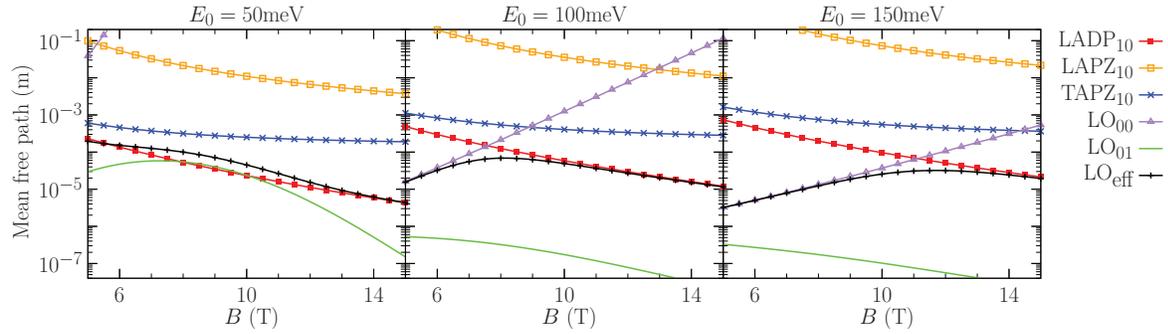}
	\caption{
		Inter-Landau-level scattering and construction of the mean-free path for the effective intra-level LO rate $\Gamma_{\rm eff}^{\rm LO}$.  Shown are the mean-free paths for $m=0$ to $m=1$ transitions due to acoustic phonon emission, along with the same for direct $m=0\to0$ and inter-level $m=1\to0$ transitions due to LO-phonon emission.
		The mean-free path of the effective LO-rate is also shown and this is determined by combining the individual rates as in Eq.~(\ref{Eq:LO_eff}).  Parameters used were a transverse confinement strength $\hbar \omega_y = 2.7$ meV, $z$-confinement length $a=3$nm.
	}
	\label{Fig:MFP_LO}
\end{figure}

Figure~\ref{Fig:MFP_LO} shows the electron mean free paths (defined as the reciprocal of the relevant decay rate multiplied by the velocity of the injected electron: $l_0 = v_0 / \Gamma$) for $m=0 \to m=1$ transitions due to acoustic-phonon emission.  We observe that these transitions are primarily driven by LADP phonons and that this scattering is most pronounced at low injection energies and high magnetic field.
This figure also shows the mean free path for the reverse transition $m=1\to m=0$ due to  LO-phonon emission (labelled LO$_{01}$), as well as the direct LO mean-free path and the effective LO mean-free path calculated from \eq{EQ:GammaLOeff}.    We see that as the injection energy increases, the effective rate becomes less dependent on the two-step inter-Landau process, especially at low magnetic field strength.
For most of the relevant range of injection energy and magnetic field, the mean free path of the  $m=1\to m=0$ transition is much shorter than that of the acoustic phonon processes. Thus, electrons are rapidly returned to the $m=0$ level, and combining rates as in \eq{EQ:GammaLOeff} is a good approximation.  However, for the lowest injection energies, the mean free paths of inter-Landau acoustic phonon scattering and the subsequent LO-phonon emission are comparable across a range of magnetic field strengths.  In this regime, there is a delay in returning the electron to the $m=0$ level and the picture provided by the effective LO rate starts to break down.  Nevertheless, for most of the parameter range considered in the subsequent, the effective rate approximation will be a good one.

\begin{figure}[t] 
	\centering
	\includegraphics[width=0.99\textwidth]{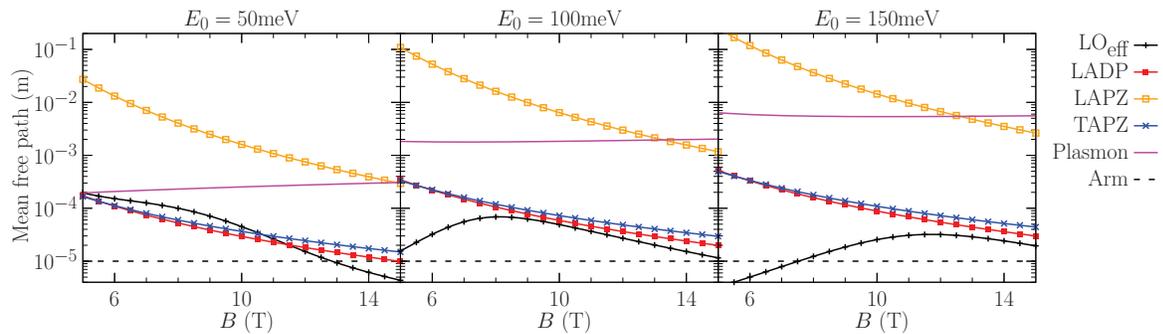}
	\caption{
		Mean free paths of hot electrons within the outermost Landau level for the indicated inelastic processes as a function of magnetic field $B$.  Results are shown for mean injection energies of $E_0=50,\,100,\,150$\,meV above the band bottom.  An arm length of 10 $\mu$m is shown for comparison. 
		We plot the mean-free paths for intra-subband scattering ($m=0 \to m=0$) for the acoustic phonons and plasmons.  The effective LO-phonon (LO$_{\rm eff}$) length is formed by combining the direct LO-scattering process with the two-step process of acoustic scattering into the $m=1$ subband followed by fast LO emission back to the outermost Landau level. 
		These results show that a balance of LO, LADP and TAPZ scattering processes are important.  Only at the very lowest magnetic field and energy does the plasmon scattering become comparable in importance with acoustic phonon scattering.
		We take a Fermi-energy of 10 meV above the band bottom, while $\omega_y$ = 2.7 meV$/\hbar$ and $a$ = 3 nm, as in Fig.~\ref{Fig:MFP_LO}.
	}
	\label{Fig:MFP}
\end{figure}

We now turn to consider scattering within the outermost Landau level, with all inter-level processes subsumed in the effective LO rate.  The mean free paths of the various processes are  shown in \fig{Fig:MFP} where they are compared with a putative MZI arm length of 10 $\mu$m.  Comparing all these processes, the effective LO mean free path is almost always the shortest, and thus this scattering channel is typically the most significant.  In the cases where it is not ($E_0 = 50$ meV panel of Fig.~\ref{Fig:MFP} and low magnetic field), the reason for this is largely the delay discussed previously within the two step LO-phonon rate.  Turning to the acoustic phonons, we see that the LAPZ interaction is always negligible, but that both LADP and TAPZ are not. In certain regimes, the rates of these becomes comparable with that of the effective LO rate.
Despite their similar mean free paths, acoustic-phonon energy relaxation is dominated by the LADP interaction because the typical energy loss per phonon is much greater than that for TAPZ phonons \cite{acoustic_phonon_theory}. 

\subsection{Electron-electron scattering}
We next discuss the effect of electron-electron interactions which, in this effectively-1D system, are dominated by the excitation of plasma modes \cite{das_Sarma,das_Sarma2,Giuliani_book}.  This can be understood by looking at the quasi-1D Coulomb interaction, $V(q) \sim K_0\left( |q \delta_\mathrm{G} | \right)$, where $K_0(x)$ is the modified Bessel function of the second kind \cite{das_Sarma2}, $q$ is the momentum exchange and $\delta_\mathrm{G}$ is the difference in guide centre of the interacting electrons.  
The direct interaction of a quasi-1D system is highly suppressed for a high energy injected electrons because the only allowed interaction of this kind is a swap of the injected quasiparticle with one within the Fermi sea, which requires $q$ to be very large.  Furthermore, the spatial separation between a hot electron and the Fermi sea is large.  As such, the interaction element $v(\gamma)$ is extremely small, resulting in a negligibly small rate.  The main source of relaxation through the Coulomb interaction is hence through plasmonic excitations.  For the high energy regime that we consider here (see Fig.~\ref{Fig:HEMZI}b), the momentum transfer $q$ is small \cite{das_Sarma_small_plas}, although the separation here is still large.  This results in a more significant interaction, though generally still weak.

In order to obtain a full solution to the rates of these processes, the inversion of the dielectric function is necessary.  The confinement we consider in this paper leads to a coupling of the $y$-axis wavefunctions to the momentum exchange in the $x$-direction.  This means we have an effectively infinite size matrix to invert in order to completely solve this problem.  We circumvent this by approximating the dielectric function as a scalar.  This is justified by the knowledge that the exchange $q$ should be small \cite{das_Sarma_small_plas}.  Hence, the range over which the interaction occurs must also be small.
Using the Hamiltonian in Eq.~(\ref{Eq:Ham_ee}) and applying Fermi's golden rule, we may obtain the decay rate due to these processes to a good approximation.  The mean free paths of \fig{Fig:MFP} 
show that relaxation due to these processes are essentially negligible for the relevant energies ($\gtrsim 50$ meV) and fields ($\gtrsim 5$ T) here.  As this process is weak, we do not expect inter-Landau level transitions to be significantly stronger, and thus we do not determine the rates for these processes.

The above results are consistent with the observations in Ref.~\cite{Phonon_Fujisawa}, where little evidence of electron-electron interactions were observed for energies above 50 meV from the band bottom. Moreover, our results suggest that the deviations from ballistic transport of electrons in Ref.~\cite{Phonon_Fujisawa} might be explained by acoustic-phonon emission rather than electron-electron interactions.

\subsection{Influence of device parameters} \label{Sec:Enhancements}

\begin{figure}[t] 
	\centering
	\includegraphics[width=0.8\columnwidth]{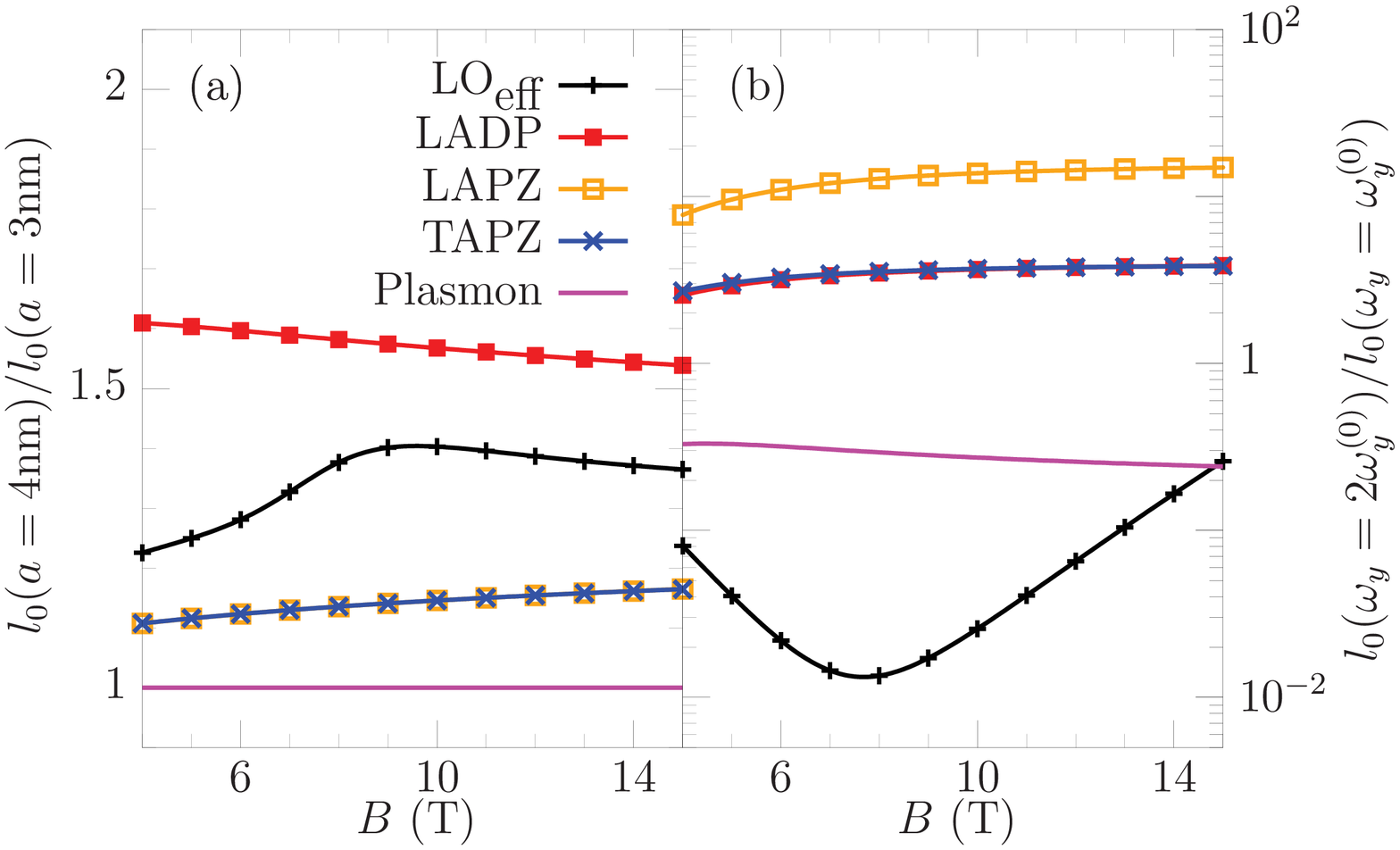}
	\caption{The mean free path of a hot electron is modified by manipulating experimentally controllable parameters.  (a) The in-plane confinement length $a$ can be modified to alter the phonon rates.  Here we plot the relative mean free path for different values of confinement length $a$. We use $a=3$ nm (used in our previous calculations) as the base value and plot for $a = 4$ nm, with $\hbar \omega_y = 2.7$ meV and $E_0 = 100$ meV.  We see that increasing $a$ is beneficial for suppressing all phonon processes, while the plasmons are left unaltered.  Although the LADP phonons are most strongly affected by this change, the relative change decreases with increasing magnetic field strength, whereas the converse is true for other acoustic phonons.  (b) Changing the transverse confinement strength $\omega_y$ creates a more complex change.  While a larger value of $\omega_y$ yields a greater mean free path for acoustic phonon processes, the direct LO phonon rate is increased, meaning a shorter mean free path.  As at large magnetic field the LO$_{\rm eff}$ rate is dominated by inter-subband acoustic phonon scattering though, the mean free path later begins increasing.  Furthermore, the mean free path due to plasmons also decreases.  We show this with the relative mean free path, again setting the base value to be that used in our previous calculations ($\omega_y^{(0)} \equiv 2.7 {\rm meV}/\hbar$) and plotting for $\omega_y = 5.4 {\rm meV}/\hbar$, while setting $a=3$ nm and $E_0 = 100$ meV.
	}
	\label{Fig:Length_ratio}
\end{figure}

The device parameters used above were chosen to match recent experiments \cite{LO_exp}.  However, the scattering rates can be altered, sometimes significantly, by these parameters, as we now discuss.
Firstly, increasing the $z$-confinement length $a$ (i.e.~weaker in-plane confinement) is known to significantly reduce the LADP phonon rates \cite{acoustic_phonon_theory}.  Increasing the value of $a$ also lowers the LO-phonon rates, whilst also suppressing LAPZ and TAPZ rates, though not by as much.  This results in decreased relaxation.  We show this in Fig.~\ref{Fig:Length_ratio}(a), where we plot the ratio of mean free path as a function of $B$ for two sample values of $a$.  We notice that the amount of suppression decreases for increasing magnetic field strength for LADP phonons, but increases for all other acoustic phonons.  The LO$_{\rm eff}$-phonon mean free path initially increases, but for larger magnetic field strengths decreases due to the dependence on inter-subband acoustic phonon scattering at these values.  Nevertheless, there is a suppression for all phonon types, at least within the magnetic field ranges that we consider here.  As such, where possible the $z$-confinement should be reduced in order to reduce scattering.  This modification can be done during the growth stage of preparing the heterostructure to be used for the device.

Moreover, the transverse confinement $\omega_y$ can also be modified by, for example, the application of top-gate potentials. The effect of changing this quantity on scattering rates is more complex.  By increasing $\omega_y$, we see an increase in the mean free paths for acoustic phonons, whereas we see a decrease for the LO$_{\rm eff}$-phonons.  Again though, for larger magnetic field this behaviour reverses when the inter-Landau process begins to dominate the direct emission, which can be seen when comparing with Fig.~\ref{Fig:MFP_LO}.  We see this in Fig.~\ref{Fig:Length_ratio}, where we consider the relative change of the mean free paths for two different confinement strengths.

It should also be mentioned that while altering the value of $a$ does not affect the plasmon rate (at least with the quasi-1D Coulomb interaction that we use here), manipulating the value of $\omega_y$ does, as can be seen from Fig.~\ref{Fig:Length_ratio}(b).  For larger values of $\omega_y$, the mean free path due to plasmon emission is decreased.  This dependence arises from the reduced separation between the hot electron and the Fermi sea for stronger transverse confinement.  In such a case, it would not necessarily be reasonable to assume that the Coulomb interaction is negligible due to the increased presence of plasmons now being induced.  Hence relaxation due to this interaction should then be considered along with phononic processes and we transition back towards the cold electron regime.

Based upon this alteration to the electron-electron interactions, it is generally inadvisable to operate at higher transverse confinement, unless operating at very high injection energies where the plasmon rate is even smaller.  In such a case, as acoustic phonons are already suppressed at high energies, it may be beneficial to increase the transverse confinement to reduce high LO-phonon rates.  Conversely, if operating at a low injection energy where acoustic phonons general dominate, it would be best to decrease the transverse confinement.  Therefore, the ideal choice of $\omega_y$ is highly dependent on the other experimental parameters to be used.


\section{Decoherence}

We now consider how the scattering processes described in Sec.~\ref{Sec:Relax} lead to decoherence within an interferometer. Based on the above, we neglect the effects of electron-electron interactions, as these processes were shown to be negligible.  With regard to acoustic phonons, as both LADP and TAPZ rates are comparable and since either process has the capacity to provide `which-way' information, we include all acoustic-phonon processes (unlike for relaxation where only LADP processes are important \cite{acoustic_phonon_theory}).

To proceed, we write down a quantum master equation for the electron density matrix $\rho$ based on rates derived from the Fr\"ohlich Hamiltonian modified to include emission within interferometer arms.
Tracking the full density matrix is unwieldy, and also unnecessary, so we therefore simplify the problem in a number of steps.
Firstly, and as in Refs.~\cite{LO_theory,acoustic_phonon_theory}, we employ a semi-classical approach to the master equation.  This means we can understand the dynamics of the system by considering only the energy-diagonal parts of the density matrix $\rho_{\alpha\beta}(E) \equiv \rho_{\alpha\beta}(E,E)$.
We assume that the wavepacket remains self-coherent upon any scattering event.  The result of this is that the wavepacket remains well localised in energy such that the decay rates are approximately constant over the wave packet.  As such, states in a wavepacket with different energies maintain their coherence throughout the evolution.  In contrast, those with the same energy but different arm index will decohere. This behaviour is coupled with a simple ballistic motion of the electron through the MZI.

Secondly, we note that electrons that have emitted an LO phonon are certainly decohered and thus contribute nothing significant to the coherence properties of the electrons at the detector. Thus, we focus only on that part of the density matrix located near the injection energy $\rho^{(0)}_{\alpha\beta}(E)$ and consider the effects of LO emission simply as an out-scattering from that portion of the density matrix.
Finally, since $\rho^{(0)}_{\alpha\beta}(E)$ is localised and its mean energy does not change greatly over the course of its transmission through the MZI, we linearise the dispersion such that we may write
$
k - k'  \approx \left[\epsilon - \left(m - m' \right) \hbar \Omega\right]/(\hbar v_0)
$
where $v_0$ is the velocity of the electron and $\epsilon$ is the energy change.

With these approximations, the master equation for the relevant part of the density matrix reads
\begin{eqnarray} \label{Eq:Master}
\dot{\rho}^{(0)}_{\alpha \beta} (E) 
& = &
-  \Gamma_{\rm LO eff} \rho^{(0)}_{\alpha \beta} (E) 
+
\int\limits_0^\infty {\rm d} \epsilon \, 
\Gamma(\epsilon) 
\left[
- \rho^{(0)}_{\alpha \beta} (E) 
+ \delta_{\alpha , \beta} \rho^{(0)}_{\alpha \alpha} (E + \epsilon) 
\right] 
\nonumber\\&&
+\int\limits_0^\infty {\rm d} \epsilon \,
\tilde{\Gamma}(\epsilon) \left[1 - \delta_{\alpha , \beta}\right] \rho^{(0)}_{\alpha \beta}(E + \epsilon) 
\, .
\end{eqnarray}
where $\Gamma (\epsilon)$ is the sum of the intra-Landau level acoustic phonon rates and $\tilde{\Gamma}(\epsilon)$ is the decay rate of the off-diagonal elements of the density matrix due to these processes.  The overall size of the density matrix describing this ensemble decreases at a rate given by $\Gamma_{\rm LO eff}$.
In obtaining the above expression we assume no band bottom.  This is reasonable due to the low energy associated with acoustic phonons and the relatively high energy of the injected electrons.
We will consider as initial state a Gaussian wave packet, of width $\sigma$ in energy space \cite{Gaussian_wavepackets,Gaussian_width}, that is fully coherent between the two arms. 
Assuming zero dispersion across the wave packet (justified since $\sigma \ll E_0$), ballistic motion of the the centre of the wavepacket along the interferometer arm gives the position as $x=v_0 t$, where $v_0=v(E_0)$ is the electron velocity at the injection energy.

The applicability of this single-subband model breaks down when the secondary LO-phonon in the two-step process is not extremely fast, as discussed in Sec.~\ref{Sec:Relax}, because without the rapid emission of the LO-phonon following the scattering into the $m=1$ Landau level, accurate tracking of populations in multiple Landau levels is required.  To avoid problems from this, we restrict ourselves to cases where the effective-rate approximation is good, and only a very small proportion of the electrons are expected to remain in the $m=1$ Landau level.  Specifically, we demand that of the subemsemble of electrons that have scattered into the $m=1$ Landau level, only at most 1\% remain there at the end of the interferometer without having emitted an LO-phonon.  In such a case, any error associated with the effective LO rate will be small and calculations using this quantity will hence be reliable.

\begin{figure}[tb] 
	\centering
	\includegraphics[width=0.8\columnwidth]{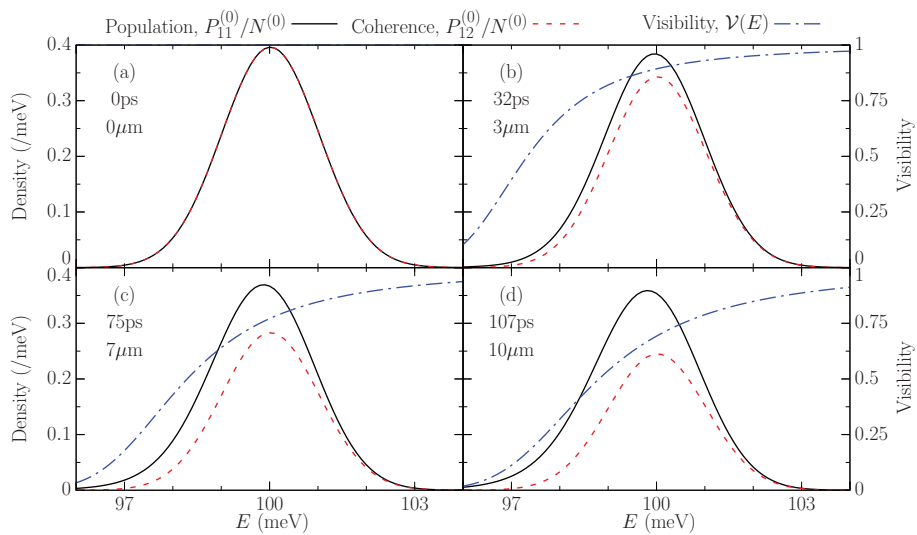}
	\caption{
		Normalised population and coherence densities, $P^{(0)}_{11}(E)/N^{(0)}$  and $P^{(0)}_{12}(E)/N^{(0)}$, as well as the energy-resolved visibility of a hot electron injected at $E_0=100$ meV as a Gaussian wave packet of width $\sigma = 1$ meV.  The normalisation removes the effects of relaxation due to LO-phonon emission.
		The four panels show results at the times indicated when the centre of the electron wave packet has travelled a distance of $x=0$, $3$, $7$, and $10$ $\mu$m from the first beamsplitter.
		At $x=0$, the electron is fully coherent and $P_{12}=P_{11}$.  But as the electron traverses the interferometer, decoherence reduces $P_{12}$ relative to $P_{11}$, and suppresses the energy-resolved visibility at the low-energy side of the wave packet.  Parameters as in \fig{Fig:MFP} with $B=12$ T.
	}
	\label{Fig:Gauss}
\end{figure}

The results of numerical evaluation of our master equation are shown in \fig{Fig:Gauss}.
The results are expressed in terms of the 
probability densities
\begin{eqnarray} \label{Eq:Prob}
P^{(0)}_{\alpha \beta}(E) 
& \equiv & \sum_k \delta \left(E - E_k \right) 
\mathrm{Tr}\left\{c_{k_\alpha}\rho^{(0)} c^\dag_{k_\beta}\right\} \, ,
\end{eqnarray}
where $c_{k_\alpha}$ is an electron annihilation operator of an edge-channel electron with wavenumber $k$ in arm $\alpha = 1,2$.  By defining these probability densities and assuming a continuum in energy space, we may track the dynamics of the system through a wavepacket like that in Fig.~\ref{Fig:Gauss}.
In the case of a symmetric interferometer with 50:50 beamsplitters, we have probability density $P^{(0)}_{11}(E) = P^{(0)}_{22}(E)$ and, with the initial conditions we consider here, the inter-arm coherence $P^{(0)}_{12}(E)=P^{(0)}_{21}(E)$ and $\mathrm{Im}\left[P^{(0)}_{12}\right]=0$.  
We also define the total weight of this portion of the density matrix as
\beq
N^{(0)} \equiv \int dE \, P^{(0)}_{11}(E) \, .
\eeq
Ignoring the effects of LO emission for a moment, Fig.~\ref{Fig:Gauss} shows that,  as the wavepacket travels, the population density $P^{(0)}_{11}(E)$ relaxes and also broadens slightly due to the stochastic nature of phonon emission \cite{acoustic_phonon_theory}.  In contrast, the coherence $P^{(0)}_{12}(E)$ shows no significant drift or diffusion, but rather shrinks in amplitude relative to  $P^{(0)}_{11}(E)$ as the wave packet propagates.

This behaviour can be understood by considering the drift-diffusion properties of the probability densities introduced in Eq.~(\ref{Eq:Prob}) with an extension of Ref.~\cite{acoustic_phonon_theory} to include inter-arm coherences.
In this approach, the elements $P^{(0)}_{\alpha\beta}(E)$ each obey a Fokker-Planck equation
\begin{eqnarray} \label{Eq:F-P}
\dot{P}^{(0)}_{\alpha\beta} & = & 
-A_{\alpha\beta} P^{(0)}_{\alpha\beta} 
+ v_{\alpha\beta}\frac{\partial P^{(0)}_{\alpha\beta}}{\partial E} 
+ D_{\alpha\beta}\frac{\partial^2 P^{(0)}_{\alpha\beta}}{\partial E^2}
.
\label{EQ:FP}
\end{eqnarray}
The coefficients here are the rate of change of amplitude, $A_{\alpha \beta}$, the drift velocity $v_{\alpha \beta}$ and the diffusion co-efficient $D_{\alpha \beta}$.  We show the expressions for these in Table \ref{Tab:F-P}, with details of the derivation given in \ref{App:F-P}.

\begin{table}
	\caption{\label{Tab:F-P} Table of Fokker-Planck co-efficients describing the evolution of the probability densities $P^{(0)}_{\alpha \beta}$.}
	\begin{indented}
		\item[]\begin{tabular}{@{}ccc}
			\br
			Coefficient & Diagonal ($\alpha=\beta$) & Off-diagonal ($\alpha \neq \beta$)\\
			\mr
			$A_{\alpha \beta}$ & $\Gamma_{\rm LOeff}$ & $\Gamma_{\rm LOeff} + \int {\rm d} \epsilon \left( \Gamma(\epsilon) - \tilde{\Gamma}(\epsilon)\right)$ \\ 
			$v_{\alpha \beta}$ & $- \int {\rm d} \epsilon \, \epsilon \Gamma(\epsilon)$ & $- \int {\rm d} \epsilon \, \epsilon \tilde{\Gamma}(\epsilon)$ \\ 
			$D_{\alpha \beta}$ & $\frac{1}{2} \int {\rm d} \epsilon \, \epsilon^2 \Gamma(\epsilon)$ &  $\frac{1}{2} \int {\rm d} \epsilon \, \epsilon^2 \tilde{\Gamma}(\epsilon)$ \\ 
			\br
		\end{tabular}
	\end{indented}
\end{table}

Central to the above analysis is that diagonal and off-diagonal elements of the density matrix have different rates, and this results in differing dynamics for the population and coherence terms.  Approximate expressions for these rates can be found using the saddle-point method, which reveals that the ratio of coherence- to population-rates is
\begin{eqnarray} \label{Eq:Rates}
\tilde{\Gamma} (\epsilon) / \Gamma (\epsilon) \approx \left[1 + \left(l_{\rm arm}/l_\Omega\right)^2\right]^{-\frac{1}{2}}
\end{eqnarray}
for all acoustic phonon types.  For realistic devices, $l_{\rm arm} \gg l_\Omega$, such that $\Gamma(\epsilon) \gg \tilde{\Gamma}(\epsilon)$. As a result, 
$v_{12} \approx D_{12} \approx 0$, while $A_{12} \approx \Gamma_{\rm LO eff} + \int {\rm d} \epsilon \, \Gamma(\epsilon)$.  Thus, the inter-arm coherence is effectively stationary in energy with an amplitude that suffers the relative decay $P^{(0)}_{12}(t) / P^{(0)}_{11}(t) \sim \exp \left(-\int {\rm d} \epsilon \, \Gamma(\epsilon)\, t \right)$.
Hence, the relative size of the coherence to the population shrinks at a rate proportional the total population relaxation rate due to intra-Landau level acoustic phonon emission.  For all realistic arm separations this behaviour will hold for almost all times, hence justifying the model chosen in Sec.~\ref{Sec:Model} of assuming constant $l_{\rm arm}$ and considering all separation into the $y$-direction.


\section{Visibility}
Assuming that the upper and lower wavepackets arrive at the second beamsplitter simultaneously \cite{Beggi,prep}, the visibility of interference fringes is directly related to the probability densities $P^{(0)}_{\alpha \beta}$.  We take the interferometric visibility to be defined as
\begin{eqnarray} \label{Eq:Vis_def}
\mathcal{V} & = & \frac{I_{\rm max} - I_{\rm min}}{I_{\rm max} + I_{\rm min}} \, ,
\end{eqnarray}
with $I_{\rm max/min}$ the maximum/minimum signal intensity measured in an output port.  In the symmetric case we find that 
the visibility at a particular energy is simply given by the ratio $\mathcal{V}(E) = P^{(0)}_{12}(E)/P^{(0)}_{11}(E)$, as we show in \ref{App:Vis}.
We initially consider the visibility of a wavepacket only subject to the decoherence mechanisms discussed in the previous section.  Figure \ref{Fig:Gauss} shows that this energy-resolved visibility is close to 1 at the high-energy end of the wavepacket, but suppressed at low-energy because the electrons here have an increased likelihood that they have undergone scattering.
To judge the interferometric performance of the interferometer, we consider the total visibility measured in an output port, which as we show in \ref{App:Vis} is given by
$
\mathcal{V}_{\rm tot} = \int {\rm d} E \, P^{(0)}_{\alpha \beta} (E)
$.

\begin{figure}[t] 
	\centering
	\includegraphics[width=0.8\columnwidth]{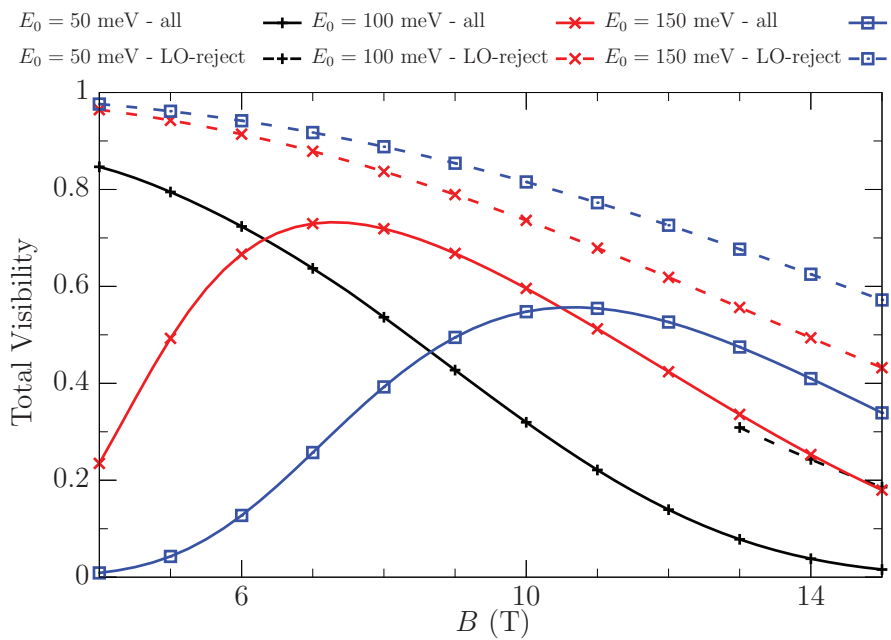}
	\caption{
		Total visibilities as a function of magnetic field strength for injection energies $E_0 = 50,100,150$ meV, with other device parameters as in Fig.~\ref{Fig:MFP}.  
		The solid lines show results for all electrons.
		Comparing with Fig.~\ref{Fig:MFP}, the total visibility peaks where the total scattering rate is minimised.
		The dashed lines show the total visibility when employing a potential barrier to block electrons that have undergone LO-phonon emission as the dashed curves.  Here we see a significant improvement at weak magnetic fields where LO-emission is dominant.  
		We include only a partial range for the filtered $E_0 = 50$ meV curve, as at lower magnetic field strengths there is a $>1\%$ proportion of electrons scattered into the $m=1$ Landau level that remain there.
	}
	\label{Fig:Tot_vis_no_filt}
\end{figure}

With the solid lines in Fig.~\ref{Fig:Tot_vis_no_filt}, we show the total visibility of all electrons as a function of magnetic field strength for select injection energies.  As might be anticipated from the relaxation rates of Fig.~\ref{Fig:MFP}, the visibility is highest in the regions of low total phonon emission.  Within these results, we see visibilities up to $\sim85\%$ are achievable.  However, with the onset of a more severe LO-phonon rate at higher energies, this visibility drops.  Moreover, where increasing magnetic field strength reduces the LO-phonon rate, the acoustic phonon rate increases, resulting in a loss of coherence and visibility through this method.  It is hence important to find a balance of these two processes, corresponding to the peaks of the solid lines in Fig.~\ref{Fig:Tot_vis_no_filt}.

\subsection{Energy filtration} \label{Sec:Filt}

To mitigate this problem and increase the visibility, we propose filtering out decoherent electrons using a potential barrier, like that shown in Fig.~\ref{Fig:HEMZI}, to block low-energy electrons.  We thus consider the detection of interference signal past the second QPC by the mean current past a potential barrier of height $E_b$ that blocks electrons with outgoing energy $E<E_b$.
We model this with a filtration function $F(E-E_b)$, centred on an energy $E_b$ where $F(x)=1$ for $x\gg0$ and $F(x)=0$ for $x\ll 0$.  In particular, electrons having undergone an LO-phonon emission are significantly lower energy and are relatively straightforward to identify and filter out.   The filters that we consider will at least block these electrons, in which case the total visibility reads (see \ref{App:Vis})
\begin{eqnarray} \label{Eq:Tot_vis_gen}
\mathcal{V}_\mathrm{tot} 
= \frac{\int_0^\infty dE \, F(E-E_b) P^{(0)}_{12}(E)}{\int_0^\infty dE \, F(E-E_b) P^{(0)}_{11}(E)} \, .
\end{eqnarray}
In the limit that the potential barrier is a sharp barrier with perfect efficiency, the filtration function becomes $F(E - E_b) = \theta(E - E_b)$, where $\theta(x)$ is the Heaviside step function, and the total visibility reads
\begin{eqnarray} \label{Eq:Tot_vis}
\mathcal{V}_{\rm tot} & = & \frac{\int_{E_b}^\infty {\rm d} E \, P^{(0)}_{1 2} (E)}{\int_{E_b}^\infty {\rm d} E \, P^{(0)}_{1 1} (E)} = \frac{1}{N^{(0)}}
\int_{E_b}^\infty {\rm d} E \, P^{(0)}_{1 2} (E) \, .
\end{eqnarray}
We initially consider a barrier height set $E_0-\hbar \omega_\mathrm{LO} + 4 \sigma \lesssim E_b \ll E_0 - 4 \sigma$ such that all (and only) LO-phonon emitted electrons are blocked.  To effectively achieve such a barrier, we must have $\sigma < \hbar \omega_{\rm LO} / 8 = 4.5$\,meV.  In this paper we assume $\sigma = 1$\,meV so satisfy this constraint, but with broader wavepackets the implementation of such barriers may not be as straightforward experimentally.  This `LO-rejection' creates greater visibilities, as we see in the dashed lines of Fig.~\ref{Fig:Tot_vis_no_filt}.  
This calculation assumes that all scattering out of the initial wavepacket region is due to LO scattering. However, as discussed previously, the breakdown of the effective-LO rate theory, particularly at low energy, means that this is not the case.  This problem is avoided by the restriction to consider cases where at most $1\%$ of the electrons scattered into the $m=1$ Landau level remain there without the emission of an LO-phonon.  This very small proportion would not have any substantial effect on our results.

\subsection{Optimisation strategies} \label{Sec:Strat}

We now propose a series of implementation schemes in order to maximise the interferometric visibility. These combine choosing better experimental parameters with various levels of energy filtration.

\begin{figure}[t] 
	\centering
	\includegraphics[width=0.8\columnwidth]{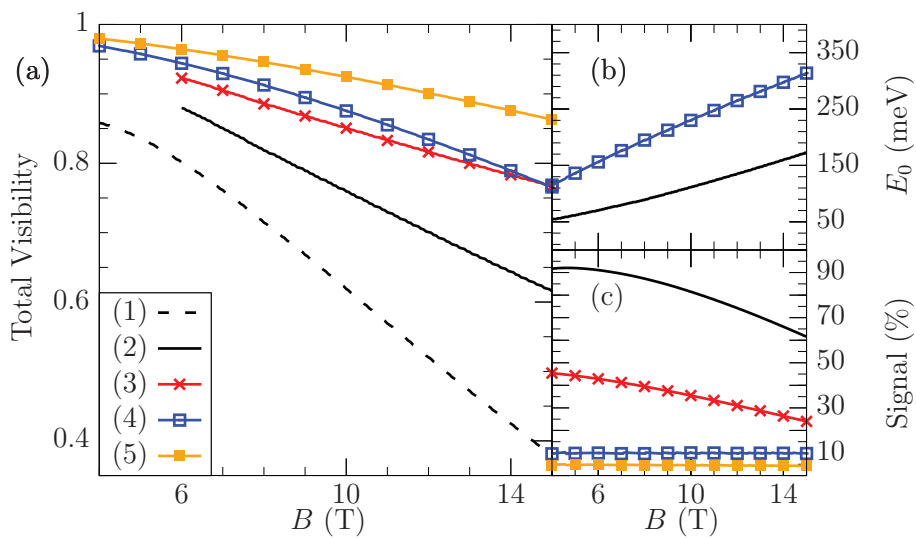}
	\caption{
		Summary of the five parameter and energy-filtration schemes. (a) Here we plot the total MZI visibility as a function of magnetic field.  We see that the total visibility decreases with increasing magnetic field, due to the increased acoustic-phonon emission rate.  With no energy filtration, as in scheme (1), the visibility decreases very rapidly.  However, when we utilise filtration techniques in schemes (2)-(5) we obtain significant improvements.  (b) Determined injection energy for the schemes.  As schemes (2) and (3), and (5) are post-processing on schemes (1) and (4) respectively, we only need to plot for two different injection energy functions.  Experimental constraints on the maximum achievable injection energy may limit the magnetic field at which these schemes can be used.  (c) The relative signal strength after applying the filtration schemes.  As expected, the schemes offering the highest visibility also require the highest amount of filtration.  Schemes (2) and (3) require lower injection energies in order to maximise the signal strength, while (4) and (5) can operate at higher energies due to the relaxed restrictions on signal loss.  The plots use the same parameters otherwise as in Fig.~\ref{Fig:MFP}, with a total arm length of 10 $\mu$m.
	}
	\label{Fig:Tot_vis}
\end{figure}

The proposal for scheme (1) is to minimise the total scattering rate from all interactions.  Here we do not implement any energy filtration and hence measure all electrons that pass through the interferometer.  In Fig.~\ref{Fig:Tot_vis}(a) we plot the total visibility as a function of the magnetic field strength.  The injection energy is also chosen as a function of the magnetic field such that the total combined scattering rate of all mechanisms is minimised.  We show this in Fig.~\ref{Fig:Tot_vis}(b).  Finding this minimum value is important as it is here that the total visibility will be maximised.  Due to the conflicting behaviour of optical and acoustic phonons, this quantity is bounded and hence it is important for experimental implementation to know what parameters to operate at.  For weak magnetic fields, we obtain good results, reaching a visibility of 85.8\%.  However, this rapidly drops off as the magnetic field is increased.

This serves as motivation to introduce the energy filtration techniques discussed in the previous subsection.  Hence, in scheme (2), we now apply LO-rejection, using the same injection energies as in scheme (1).  As in Fig.~\ref{Fig:Tot_vis_no_filt} we see an improvement in the visibility of this signal.  We do not show results for low values of magnetic field for this scheme due to the reliability of the effective LO-rate.  Obtaining this higher visibility is at a cost of a reduced signal strength, which is shown in Fig.~\ref{Fig:Tot_vis}(c).  The signal strength is high at low magnetic field, but decreases due to the increased effective LO-phonon rates at high magnetic fields and injection energy that can be seen in Fig.~\ref{Fig:MFP}.

In our next scheme, we place the potential barrier at $E_b = E_0$ and operate at the same injection energies as before.  As observed in Fig.~\ref{Fig:Gauss}, the energy-resolved visibility is lower for lower energies within the wavepacket due to an increased likelihood of scattering having occurred.  By placing the potential barrier in this way and hence implementing `wavepacket filtration', we cut off a part that is more likely to have scattered and hence become decoherent.  We hence see in scheme (3) a further increase in visibility.  This is of course at a cost of further signal loss, as we now remove around half of the already LO-filtered wavepacket.

In scheme (4) we revert to the LO-rejection method of the potential barrier but choose to accept higher loss of signal.  This allows us to operate at higher energies where acoustic phonons are more suppressed.  The results in \fig{Fig:Tot_vis} are obtained with signal loss of 90\%, which increases the visibility up to 96.9\% at 4 T.  This scheme begins to show limitations at the high end of our magnetic field range, where the total visibility is similar to that in scheme (3), but with a more severe signal loss.  This demonstrates that at high magnetic field strengths it is acoustic phonons that are the significant decoherence mechanism and hence require extra attention.

With this point in mind, the final scheme (5) is a combination of schemes (3) and (4), where we use a potential barrier to create wavepacket filtration and also choose an injection energy to give a LO loss of 90\%. This gives a very high visibility, up to 98.0\%, but results in a signal with around 95\% loss.  While there is only a small difference in the best visibilities from this scheme with (3), it performs significantly better at high $B$ ($\sim 86\%$ rather than $\sim 77\%$ at $B=15$ T).  This is because we now filter some of the electrons having undergone acoustic phonon decoherence as well as the optical phonon processes dealt with by scheme (4).

\section{Conclusions}
In this paper we have presented an overview of the mechanisms by which hot electrons from dynamical quantum dot charge pumps undergo relaxation and decoherence.
Electron-electron interactions play only a very small role in hot-electron regime and therefore do not limit the coherence of hot-electron devices.  It is instead the emission of phonons that lead to decoherence.  Through our analysis, we have determined the rates at which these occur for all phonon types.  In particular, we find that LO-phonon emission is generally the dominant mechanism for this, whether it is through direct emission or through a two-step inter-Landau process \cite{acoustic_phonon_theory,LO_exp}.

Concerning the MZI, we have shown how to maximise the visibility by determining the optimum injection energies as a function of magnetic field to reduce scattering events.  In doing so, we predict visibilities $\gtrsim 85\%$ over a relatively long distance of $10 \mu$m for typical confinement geometries.
One further way in which the visibility might be enhanced is to modify the confinement of the electrons.  In particular, decreasing the $z$-confinement decreases phonon rates \cite{acoustic_phonon_theory} and should also improve visibility.  Increasing the in-plane transverse confinement $\omega_y$ has a more complicated effect but generally, it further increases visibility at the expense of greater signal loss from LO-phonon emission.  This must be carefully balanced with the onset of electron-electron interactions however, as too much of an increase will make this interaction non-negligible.

By using potential barriers to filter out lower-energy electrons, we have devised a number of further schemes to obtain even greater visibilities.  These schemes allow us to sacrifice signal strength for visibility in various ways, and we have shown that visibilities of $\sim 98\%$ over 10 $\mu$m are achievable in this way. 
This filtration works because the primary decoherence mechanism here (phonon emission) leaves its trace on the energy of the electrons.  By employing an energy filter, we remove from the ensemble those electrons that are most likely to have undergone decoherence.   This in turn means that -- based on the independent criterion of energy -- we are able to extract a subensemble with a higher degree of coherence at the detector than is possessed by the ensemble overall.
As an alternative strategy to energy filtration, a similar effect could potentially be achieved by means of a time gate, since the lower energy electrons will also have lower velocities and thus arrive later at the detector. This method may not be as effective as energy filtration, however, since the time-delay effects may be small compared with the ramp-up time of the detector barriers.

We note that filtration does not reduce the total amount of decoherence, but rather selects those electrons that possess higher levels of coherence.  As such, the total information content of the interference effects within the signal will not be enhanced, but it will provide a signal more suitable for further quantum tasks and processing.
While undesirable, a high signal loss may be acceptable in hot-electron quantum optics due to the high rate of operation of the electron pumps \cite{Charge_pump}, meaning the loss of even a large proportion of the injected electrons would still produce strong signals.

It should also be acknowledged that we have based these results on a MZI with arm length of $10 \mu$m, which is a greater distance than has been considered previously \cite{eMZI,eMZI_exp,eMZI_decoherence2,eMZI_exp2}.  Despite this, we still predict a significantly higher visibility than those previously observed.
These results therefore indicate the potential of hot-electrons as a platform for coherent electronic devices, as well as providing insight into the optimal conditions in which to conduct such experiments.

\section*{Acknowledgements}
This research was supported by EPSRC Grant No. EP/P034012/1. M.K. was supported by the UK Department for Business, Energy, and Industrial Strategy, and by 17FUN04 SEQUOIA. This project has received funding from the European Metrology Programme for Innovation and Research (EMPIR) program, cofinanced by the Participating States, and from the European Union’s Horizon 2020 Research and Innovation Program.

\appendix

\section{Determination of phonon decay rates} \label{App:Rates}
Beginning with Eq.~(\ref{Eq:Ham}), we use Fermi's golden rule to determine the relevant decay rates for phonon scattering events.  The matrix element $\Lambda_{{\bf n} {\bf n'}}^{(\gamma)}$ is defined as
\begin{eqnarray}
\Lambda_{{\bf n} {\bf n'}}^{(\gamma)} ({\bf q}) & = & M^{(\gamma)} \left({\bf q} \right) \delta_{q_x , k' - k} G^{(y)}_{m' k' , m k} (q_y) G_{n' n}^{(z)} (q_z) \, ,
\end{eqnarray}
where $G^{(y)}_{m' k' , m k}$ and $G_{n' n}^{(z)}$ are overlap functions
\begin{eqnarray}
G^{(y)}_{m' k',mk} (q_y) & = & \int {\rm d}y \, {\rm e}^{{\rm i} q_y y} \chi^*_{m' k'} (y) \chi_{m k} (y) \nonumber \\
G^{(z)}_{11} (q_z) & = & \int {\rm d} z \, \ {\rm e}^{{\rm i} q_z z} \phi^* (z) \phi(z) \, ,
\end{eqnarray}
and $M^{(\gamma)} ({\bf q})$ is the matrix element of phonon type $\gamma$ \cite{acoustic_phonon_theory}, whose forms are \cite{Mahan_book,Climente}
\begin{eqnarray}
|M^{\rm LO} ({\bf q})|^2 & = & \frac{4 \pi \alpha \hbar \left(\hbar \omega_{\rm LO}\right)^{\frac{3}{2}}}{{\left(2 m_e^*\right)}^{\frac{1}{2}} V} \frac{1}{\left| {\bf q} \right|^2} \\ \left| M^{\rm LADP} \left({\bf q} \right) \right|^2 & = & \frac{\hbar D^2}{2 d c_{\rm LA} V} \left| {\bf q} \right| \nonumber \\
\left| M^{\rm LAPZ} \left({\bf q} \right) \right|^2 & = & \frac{32 \pi^2 \hbar e^2 h_{14}^2}{\epsilon_r^2 d c_{\rm LA} V} \frac{\left(3 q_x q_y q_z\right)^2}{\left| {\bf q} \right|^7} \nonumber \\ \left| M^{\rm TAPZ} \left({\bf q} \right) \right|^2 & = & \frac{32 \pi^2 \hbar e^2 h_{14}^2}{\epsilon_r^2 d c_{\rm TA} V} \left| \frac{q_x^2 q_y^2 + q_y^2 q_z^2 + q_z^2 q_x^2}{\left| {\bf q} \right|^5} - \frac{\left(3 q_x q_y q_z\right)^2}{\left| {\bf q} \right|^7} \right| \, , \nonumber
\end{eqnarray}
with volume element $V$, electron effective mass $m_e^* = 0.067 m_e$,  Fr\"ohlich coupling constant $\alpha = 0.068$, density $d = 5310$ kg m$^{-3}$ and $D = 8.6$ eV is the acoustic deformation potential.  Meanwhile, $h_{14} = 1.41 \times 10^9$ V/m, $\epsilon_r = 12.9$ and $c_{\rm LA} = 4720$ ms$^{-1}$ and $c_{\rm TA} = 3340$ ms$^{-1}$ are the respective phonon velocities.

The relaxation due to these decay rates is already well understood, both for LO-phonons \cite{LO_theory} and acoustic \cite{acoustic_phonon_theory} and hence we do not explain this further here.  However, we then consider a full quantum picture to derive a master equation for the interferometer.  Here, we consider intra-Landau level scattering through acoustic phonons.  Electrons that have undergone other forms of scattering (LO-phonons or acoustic inter-Landau level scattering) are trivially decoherent and are highly distinguishable from the injected electron.  However, intra-Landau level acoustic phonon scattering has a more subtle effect, as we shall now see.

The decay rate coming into the calculations here is
\begin{eqnarray}
\Gamma_{{\bf n'}_\alpha {\bf n}_\beta}^{(\gamma)} & = & \sum\limits_{\bf q} \Lambda^{(\gamma)}_{{\bf n}_\alpha {\bf n'}_\beta} \Lambda^{(\gamma)*}_{{\bf n}_\beta {\bf n'}_\beta} \delta \left(E_{\bf n'} - E_{\bf n} + \hbar \omega \right) \exp \left[ {\rm i} q_y l \left(\delta_{\beta , 2} - \delta_{\alpha, 2}\right)\right] \, . 
\end{eqnarray}
For $\alpha \neq \beta$, a phase difference in the phonon field is accumulated due to the arm separation $l_{\rm arm}$.  As stated in the main body, we place all dependence on this perpendicular to the propagation direction.  This is due to that only for small $l_{\rm arm} \sim l_\Omega$ will this rate be significant.

Now transforming to polar co-ordinates and taking the continuum limit for ${\bf q}$, the rate becomes
\begin{eqnarray}
\frac{L}{2 \pi} \Gamma_{{\bf n'}_\alpha {\bf n}_\beta}^{(\gamma)} & = & \frac{2 \pi}{\hbar} \left(\frac{L}{2 \pi}\right)^3 \frac{1}{\hbar c_{\rm A}} \int\limits_0^\infty {\rm d} q \, \int\limits_1^{-1} {\rm d}\left(\cos\left(\theta\right)\right) \, \int\limits_0^{2 \pi} {\rm d} \phi \nonumber \\
&& \times \delta\left(q - q_0\right) \delta\left(\cos(\theta) - \frac{k' - k}{q} \right) \left|M^{(\gamma)} \left(q,\theta,\phi\right) \right|^2 \nonumber \\
&& \times \left|G^{(y)}_{m' k' , m k} \left(q \sin(\theta) \cos(\phi) \right) \right|^2 \left|G^{(z)} \left(q \sin(\theta) \sin(\phi)\right)\right|^2 \nonumber \\
&& \times \exp\left({\rm i} q \sin(\theta) \cos(\phi) l \left(\delta_{\beta , 2} - \delta_{\alpha , 2} \right) \right) \, ,
\end{eqnarray}
with $c_{\rm A} = c_{\rm LA},c_{\rm TA}$ depending on the phonon type, $L$ a quantisation length such that $L^3 = V$ and $q_0 = \left(E_{\bf n} - E_{\bf n'}\right) / \hbar c_A$.

We see there are two different rates present, namely $\Gamma \equiv \Gamma_{\alpha = \beta}$ and $\tilde{\Gamma} \equiv \Gamma_{\alpha \neq \beta}$.  For the first case, we obtain the standard decay rates present in relaxation processes previously studied \cite{acoustic_phonon_theory}.  However, in the off-diagonal case, corresponding to inter-arm coherences, there is an extra damping term.

\section{Fokker-Planck equation} \label{App:F-P}

We may find a solution to Eq.~(\ref{Eq:Master}) by using a cumulant generating function approach.  Defining
\begin{eqnarray}
\rho^{(0)}_{\alpha \beta} (\chi,t) & \equiv & \int\limits_{-\infty}^\infty {\rm d} E \exp\left({\rm i} \chi E\right) \rho^{(0)}_{\alpha \beta} \left(E,t\right) \, ,
\end{eqnarray}
we may use this in Eq.~(\ref{Eq:Master}) to get
\begin{eqnarray}
\dot{\rho}^{(0)}_{\alpha \beta} \left(\chi,t\right) & = & 
\left[
-\Gamma_{\rm LOeff}
+\delta_{\alpha , \beta} \Lambda(\chi) + \left(1 - \delta_{\alpha , \beta} \right) \tilde{\Lambda}(\chi) - \Lambda(0) 
\right] \rho^{(0)}_{\alpha \beta} (\chi,t) \, , \nonumber \\
\end{eqnarray}
where we introduce the terms
\begin{eqnarray}
\Lambda(\chi) & \equiv & \int\limits_0^\infty {\rm d} \epsilon \, \Gamma (\epsilon) \exp\left(- {\rm i} \chi \epsilon \right) \nonumber \\
\tilde{\Lambda}(\chi) & \equiv & \int\limits_0^\infty {\rm d} \epsilon \, \tilde{\Gamma} (\epsilon) \exp\left(- {\rm i} \chi \epsilon \right) \, .
\end{eqnarray}
With these definitions, the corresponding cumulant generating function is
\begin{eqnarray}
{\mathcal{F}}_{\alpha \beta} (\chi,t) & = & \left(-\Gamma_{\rm LOeff}
+\delta_{\alpha , \beta} \Lambda(\chi) + \left(1 - \delta_{\alpha , \beta} \right) \tilde{\Lambda}(\chi) - \Lambda (0) \right) t + {\mathcal{F}}_{\alpha \beta} (\chi,0) \, .  \nonumber \\
\end{eqnarray}
From this, the relevant energy cumulants can be obtained via
\begin{eqnarray} \label{Eq:Cum}
\left<E^s (t) \right>_{\alpha \beta , {\rm c}} & = & \left. \frac{\partial^s}{\partial \left({\rm i} \chi\right)^s} {\mathcal{F}}_{\alpha \beta}\left(\chi,t\right) \right|_{\chi = 0} \, .
\end{eqnarray}
Using these energy cumulants, we may construct a Fokker-Planck master equation describing the probability densities of a particle being found with energy $E$, as introduced in Eq.~(\ref{Eq:F-P}).  The rate of change of amplitude $A_{\alpha \beta}$, drift-velocity $v_{\alpha \beta}$ and diffusion parameter $D_{\alpha \beta}$ can be found by taking the $s=0,1,2$ solutions of Eq.~(\ref{Eq:Cum}) respectively, the results of which are summarised in Table \ref{Tab:F-P}.

\section{Interferometer visibility} \label{App:Vis}

In order to quantify the sensitivity of the interferometer, we consider the visibility, starting with the definition given in Eq.~(\ref{Eq:Vis_def}).  Suppose a single particle of energy $E$ is injected into an interferometer, constructed with a pair of identical 50:50 beamsplitters, with density matrix $\rho = \ket{1}\bra{1}$, with $\ket{\alpha}$ corresponding to a particle in arm $\alpha$.  After passing through the first beamsplitter and traversing the length of the interferometer before arriving at the second beamsplitter, the density matrix describing the particle is
\begin{eqnarray}
\rho & = & \frac{1}{2} \left( \ket{1}\bra{1} + \varepsilon {\rm e}^{-{\rm i} \varphi} \ket{1}\bra{2} +\varepsilon {\rm e}^{{\rm i} \varphi} \ket{2} \bra{1} + \ket{2} \bra{2} \right) \, , \nonumber \\
\end{eqnarray}
where $\varphi$ is the phase of arm 2 relative to arm 1 and $\epsilon$ describes the amount of decoherence between the two arms that has occurred ($\varepsilon = 1$ for no decoherence, $\varepsilon = 0$ for complete decoherence).  Accordingly, $\varepsilon$ can be expressed as $\varepsilon = \rho_{\alpha \beta} / \rho_{\alpha \alpha}$ for $\{\alpha , \beta\} \in [1,2]$.

After passing through the second beamsplitter, the density matrix becomes
\begin{eqnarray} \label{Eq:MZI_end}
\rho & = & \frac{1}{2} \left(1 - \varepsilon \cos(\varphi) \right) \ket{1}\bra{1} - \frac{{\rm i} \varepsilon}{2} \sin(\varphi) \ket{1}\bra{2} \\
&& + \frac{{\rm i} \varepsilon}{2} \sin(\varphi) \ket{2} \bra{1} + \frac{1}{2} \left(1 + \varepsilon \cos(\varphi) \right) \ket{2}\bra{2} \, . \nonumber 
\end{eqnarray}
The maximum and minimum intensities can be determined by looking at the diagonal elements of Eq.~(\ref{Eq:MZI_end}).  Computing this and substituting into Eq.~(\ref{Eq:Vis_def}), we find that $\mathcal{V} = \varepsilon$ for both arms in an interferometer with 50:50 beamsplitters.

Considering an electron like those that we consider in this paper, the particle does not have a definite energy and, as the phonon emission rates are energy dependent, the decoherence parameter will also become energy dependent.  However, for each value of $E$ the energy resolved visibility will be
\begin{eqnarray}
\mathcal{V} (E) & = & \varepsilon(E) = \frac{P^{(0)}_{\alpha \beta} (E)}{P^{(0)}_{\alpha \alpha} (E)} \, ,
\end{eqnarray}
for those electrons that have not emitted an LO phonon, and $\mathcal{V} (E) =0$ for those that have.

The total visibility $\mathcal{V}_{\rm tot}$, which we define as the weighted sum of all the energy-resolved visibilities with the corresponding population density at that point is then
\begin{eqnarray}
\mathcal{V}_{\rm tot} & = & \int {\rm d} E \, P^{(0)}_{\alpha \alpha} (E) \mathcal{V}(E) = \int {\rm d} E \, P^{(0)}_{\alpha \beta} (E) \, .
\end{eqnarray}
In the proposed optimisation schemes, we cut off part of the injected Gaussian wavepacket by placing a potential barrier of height $E_b$ in order to maximise the total visibility of the signal.  In such a case, the calculation of the total visibility requires a renormalisation based upon the probability of being at such energy.  Accordingly, the total visibility becomes
\begin{eqnarray} \label{EQ:VEb}
\mathcal{V}_{\rm tot} (E_b) & = & \frac{\int_{0}^\infty {\rm d} E \, F(E - E_b) P_{\alpha \alpha} (E) \mathcal{V}(E)}{\int_{0}^\infty {\rm d} E \, F(E - E_b) P_{\alpha \alpha} (E)} 
\end{eqnarray}
where $P_{\alpha \alpha} (E)$ is the complete probability density (not just the upper part) and $F(E - E_b)$ is an increasing function in the interval $[0,1]$ that acts as a filtration function due to the potential barrier. If the filtration function is chosen such that it reliably blocks the portion of the spectrum corresponding to LO phonons having been emitted, \eq{EQ:VEb} becomes
\begin{eqnarray} \label{EQ:VEb2}
\mathcal{V}_{\rm tot} (E_b) 
& = & 
\frac{
	\int_{0}^\infty {\rm d} E \, F(E - E_b) P^{(0)}_{\alpha \alpha} (E) \mathcal{V}(E)
}
{
	\int_{0}^\infty {\rm d} E \, F(E - E_b) P^{(0)}_{\alpha \alpha} (E)
} 
\nonumber \\
& = & 
\frac{\int_{0}^\infty {\rm d} E \, F(E - E_b) P^{(0)}_{\alpha \beta} (E)}{\int_{0}^\infty {\rm d} E \, F(E - E_b) P^{(0)}_{\alpha \alpha} (E)} \, ,
\end{eqnarray}

\newpage
\begin{center}
	{\bf References}
\end{center}
\bibliography{references}

\end{document}